\documentstyle[11pt,twoside]{article}

\input{epsf.sty}

\newcommand\PrOsSb{PrOs$_4$Sb$_{12}$}

\begin{document}

%Title of paper
\title{Electrical resistivity and magnetization measurements on the heavy
       fermion superconductor \PrOsSb{}}

\author{P.-C. Ho\thanks{pch@physics.ucsd.edu},
        N. A. Frederick, V. S. Zapf, E. D. Bauer, T. D. Do, and M. B. Maple
        \\Institute for Pure and Applied Physical Sciences
        \\and Department of Physics
        \\University of California, San Diego
        \\La Jolla, CA 92093-0360, U.S.A.\\
        \\A. D. Christianson and A. H. Lacerda
        \\National High Magnetic Field Laboratory/LANL
        \\Los Alamos, NM 87545}

\date{March 19, 2003}

%\maketitle must follow title, authors.
\maketitle

\begin{abstract}
The filled skutterudite compound \PrOsSb{}, the first example of a
Pr-based heavy fermion superconductor, displays superconductivity
with $T_c\sim 1.85$ K and has an effective mass $m^* \sim$
\mbox{50 $m_e$}, where $m_e$ is the free electron mass. For
magnetic fields above 4.5 T, sharp features in the normal state
electrical resistivity, magnetization, specific heat, and thermal
expansion data suggest the occurrence of a phase transition at
high fields. This high field ordered phase in the normal state may
originate from a combination of crystalline electric field
enhanced Zeeman splitting and quadrupolar ordering. We present an
investigation of the electrical resistivity and magnetization of
\PrOsSb{} as a function of temperature between \mbox{350 mK} and
\mbox{3.5 K} and magnetic field up to \mbox{18 T}. The data reveal
a detailed phase boundary of the high field ordered phase as well
as the lower critical field $H_{c1}$ and the onset field of the
peak effect in the superconducting state of \PrOsSb{}.


keywords: \PrOsSb{}, magnetoresistance, magnetization, peak
          effect, $H_{c1}$, high field ordered phase
\end{abstract}

% body of paper here - Use proper section commands
% References should be done using the \cite, \ref, and \label commands
\section{Introduction}
The filled skutterudite compound \PrOsSb{} displays
superconductivity with $T_c\approx$ \mbox{1.85 K} and has an
effective mass $m^*\sim$ \mbox{50 $m_e$}.~\cite{Bauer02} This
compound is the first example of a Pr-based heavy fermion
superconductor (all other known heavy fermion superconductors are
intermetallic compounds of the rare earth element Ce or the
actinide element U). Inelastic neutron scattering experiments,
along with an analysis of magnetic susceptibility $\chi(T)$ and
specific heat $C(T$) data~\cite{Maple02a,Maple02b} for a cubic
crystalline electric field (CEF), yield a Pr$^{3+}$ energy scheme
consisting of a nonmagnetic $\Gamma_3$ doublet ground state
(\mbox{0 K}), a $\Gamma_5$ triplet first exited state
(\mbox{$\sim$ 8 K}), and higher energy $\Gamma_4$ triplet
(\mbox{$\sim$ 133 K}) and $\Gamma_1$ singlet (\mbox{$\sim$ 320 K})
excited states. The heavy fermion properties of the Pr-based
compounds PrInAg$_2$ and PrFe$_4$P$_{12}$ have been attributed to
the interaction of the charges of the conduction electrons with
the electric quadrupole moments of the Pr$^{3+}$ $\Gamma_3$
nonmagnetic doublet ground state in the
CEF.~\cite{Yatskar96,Sato00} The evidence for a Pr$^{3+}$
$\Gamma_3$ ground state in \PrOsSb{} indicates that the electric
quadrupolar fluctuations may be responsible for the heavy fermion
state in this compound and could also be involved in the
superconductivity.

For magnetic fields $H$ above $\sim 4.5$ T, sharp features in
measurements of the normal state electrical
resistivity~\cite{Ho02} $\rho(T)$, magnetization~\cite{Tenya02}
$M(H)$, specific heat~\cite{Vollmer03} $C(T)$ and thermal
expansion~\cite{Oeschler02} $\alpha(T)$ of \PrOsSb{} indicate that
a phase transition is induced at high fields. The origin of the
high-field ordered phase (HFOP) is still under investigation but
may be related to the crossing of the Zeeman levels of the
$\Gamma_3$ and $\Gamma_5$ CEF states and a corresponding change of
the ground state at high fields.~\cite{Vollmer03} In this report,
we present further results of our investigation of the electrical
resistivity measurements up to \mbox{18 T} and magnetization
measurements with magnetic field $H \parallel [111]$ and [001]
crystallographic directions up to \mbox{5.5 T}.

\section{Experimental Details}
The \PrOsSb{} samples studied were single crystals grown in Sb
flux.~\cite{Bauer02b} X-ray diffraction measurements confirmed the
cubic LaFe$_4$P$_{12}$-type structure. The $\rho(H,T)$
measurements were made with a Linear Research LR 700 4-wire ac
bridge operating at 16 Hz with constant currents of \mbox{100
$\mu$A} (\mbox{0 T} $\le H \le$ \mbox{10 T}) and \mbox{300 $\mu$A}
(\mbox{10 T} $\le H \le$ \mbox{18 T}), and in a transverse
geometry in a $^3$He-$^4$He dilution refrigerator~\cite{Ho02}
(\mbox{0 T} $\le H \le$ \mbox{10 T}) and a $^3$He cryostat
(\mbox{10 T} $\le H \le$ \mbox{18 T}). The $M(H,T)$ measurements
were performed in a Faraday magnetometer in a $^3$He refrigerator
with a gradient field of \mbox{1 kOe/cm}.

\section{Results and Discussion}
\begin{figure}
 %\begin{center}
  \epsfbox{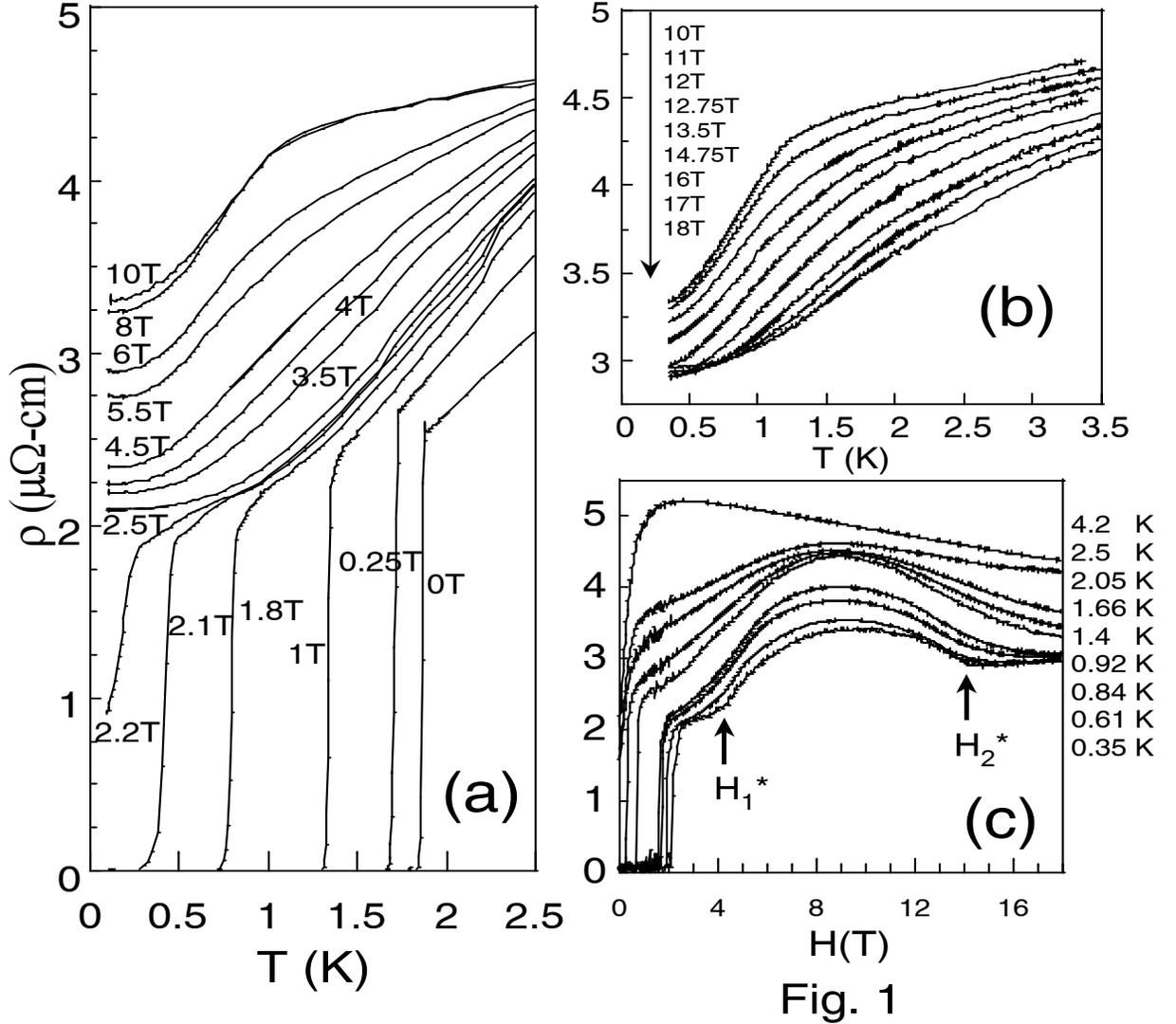}
 %\end{center}
 \caption{(a)(ref. 6) and (b) Electrical resistivity $\rho$ vs $T$
          in various magnetic fields $H$ up to 18 T for \PrOsSb{}
          single crystals. (c) $\rho$ vs $H$ at various $T$ up to 4.2 K.
          The rapid drop in $\rho$ to zero for $H <$ 2.5 T is due to the
          superconducting transition, while the shoulder in
          $\rho(T)$ at $\sim$ 1 K above 4.5 T and sharp kinks in
          $\rho(H)$ (marked as $H_1^*$ and $H_2^*$) below 0.7 K are
          due to a field induced phase
          (high field ordered phase).~\cite{Vollmer03}}
 \label{fig:rhoT&H}
 \end{figure}

\begin{figure}
 %\begin{center}
  \epsfbox{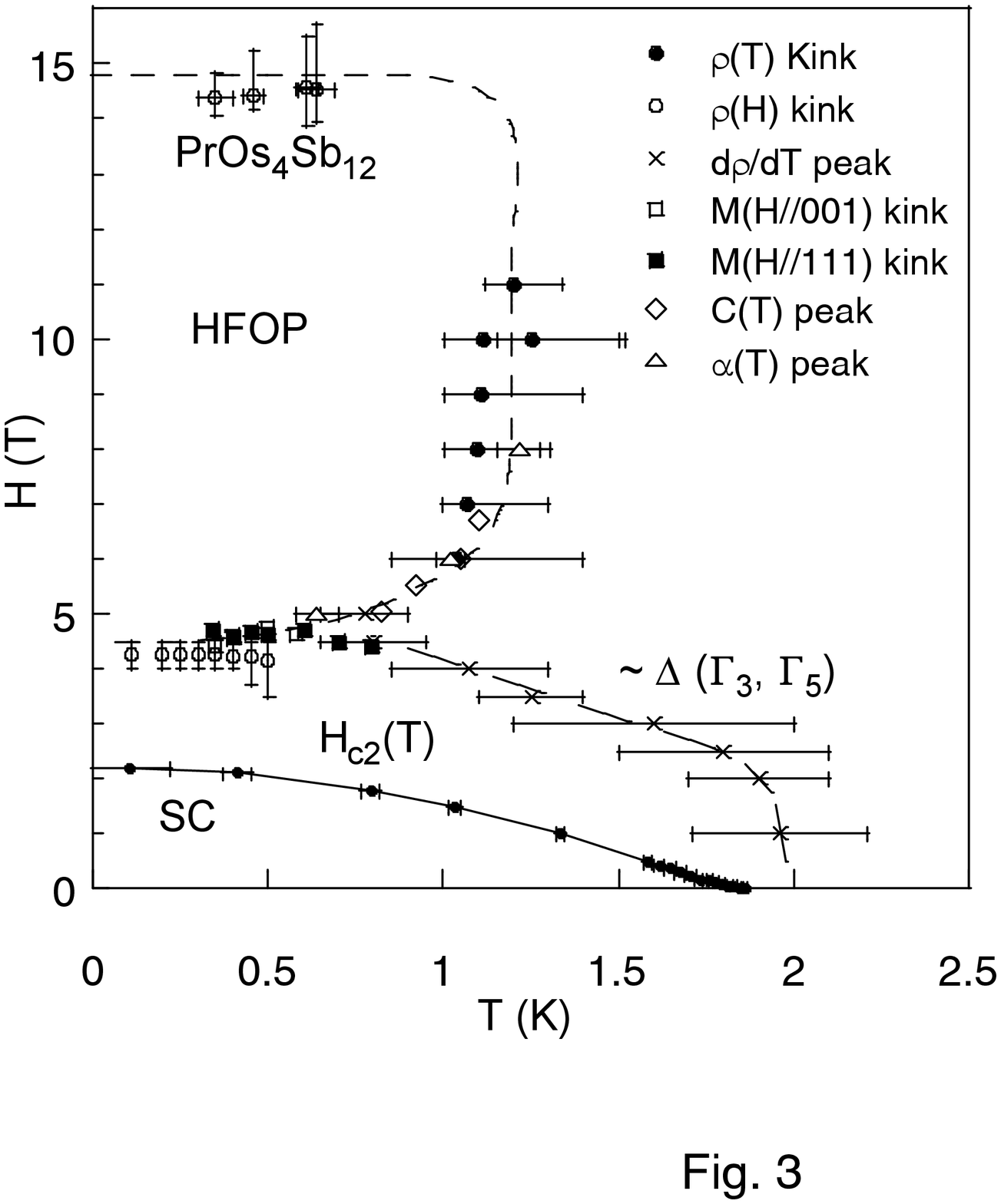}
  %\end{center}
 \caption{$H-T$ phase diagram for \PrOsSb{}. The superconducting state
          (SC) phase boundary is derived from the electrical resistivity
          $\rho(T,H)$ data.~\cite{Bauer02,Ho02}
          The high field ordered phase (HFOP) is deduced from the
          features observed in $\rho(T,H)$,
          $C(T,H)$,~\cite{Vollmer03} $M(H,T)$ ($H \parallel$ [001] and [111]),
          and $\alpha(T,H)$.~\cite{Oeschler02} The dashed line,
          drawing through the points where $d\rho/dT$ exhibits a peak above 1 K,
          is a measure of the energy difference between the Pr$^{3+}$ $\Gamma_3$
          ground and $\Gamma_5$ first excited state.~\cite{Vollmer03,Maple02b}}
 \label{fig:PhaseDiagram}
\end{figure}

%rhoT and rhoH factor of 0.4445
The behavior of electrical resistivity $\rho$ below \mbox{4.2 K}
and \mbox{18 T} is summarized in Fig.~\ref{fig:rhoT&H}.
Figs.~\ref{fig:rhoT&H}(a)[ref.~6] and (b) show $\rho(T)$ data in
various constant fields and Fig.~\ref{fig:rhoT&H}(c) shows
isotherms of $\rho(H)$. In this temperature range, the phonon
contribution to $\rho$ is negligible.~\cite{LaOs4Sb12} The
\mbox{$\rho (T,H)$} data in Figs.~\ref{fig:rhoT&H}(b) and (c) were
measured on a different \PrOsSb{} sample and in a different
cryostat. The absolute value of the resistivity is
sample-dependent in the present work, probably due to irregular
sample shapes and the presence of microcracks in the samples. The
$\rho(T,H)$ data in Figs.~\ref{fig:rhoT&H}(b) and (c) are
normalized by comparing the $\rho(T)$ data at 10 T for both
samples, which differ by a factor of $\sim 0.44$. The sharp drops
in $\rho(T)$ below \mbox{2.5 T} are due to the superconducting
transition. Above $\sim$ \mbox{4.5 T}, a kink starts to develop in
the $\rho(T)$ curves and becomes most pronounced between \mbox{7
T} and \mbox{11 T}, then gradually subsides as $H$ increases. This
kink is apparently related to the occurrence of a high field
ordered phase which will be discussed later
(Fig.~\ref{fig:PhaseDiagram}). The $\rho(T)$ data between \mbox{8
T} and \mbox{10 T} almost overlap with each other
(Fig.~\ref{fig:rhoT&H}(a) and ref. 6). The $\rho(H)$ data below 1
K reach a maximum at $\sim$ \mbox{9 T} as shown in
Fig.~\ref{fig:rhoT&H}(c), while the dome-shaped feature becomes
more pronounced as $T$ decreases. Two sharp kinks ($H_1^*$ and
$H_2^*$) become easily identified below \mbox{0.61 K} and mark the
boundary of the high field ordered
phase.~\cite{Ho02,Tenya02,Vollmer03}

\begin{figure}
 %\begin{center}
  \epsfbox{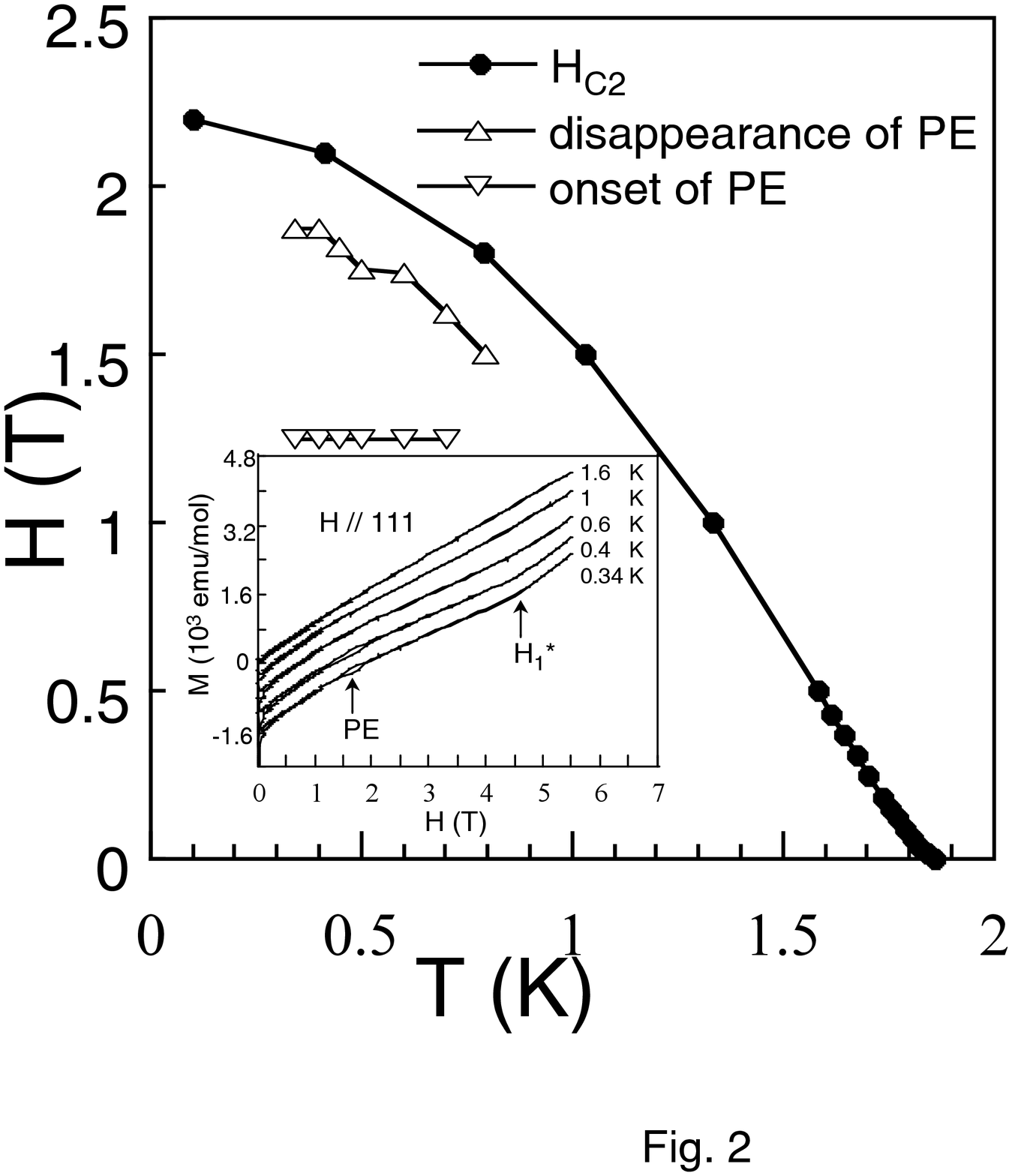}
 %\end{center}
 \caption{Onset and disappearance fields of the peak effect (PE)
          in the $H-T$ phase diagram of a \PrOsSb{} single crystal with
          $H \parallel [111]$. $H_{c2}$ is derived from the superconducting
          transition observed in the $\rho(T)$ data.~\cite{Bauer02,Ho02}
          Inset: $M$ vs $H$ ($\parallel [111]$) at various
          temperatures. The $M(H)$ kink occurs at $H_1^*$ where the
          boundary of the high field ordered phase lies.}
 \label{fig:PEH111}
\end{figure}

%PE
The magnetization $M(H)$ measurements were performed with the
applied magnetic field oriented along the [111] and [001]
crystallographic directions of a single crystal of \PrOsSb{} for 0
T $\le H\le$ 5.5 T. The inset of Fig.~\ref{fig:PEH111} only shows
the isothermal $M(H)$ data for $H\parallel [111]$. A kink (at
$H_1^*$) appears in $M(H)$ curves above 4.5 T and can not be
resolved above 0.8 K for $H\le$ 5.5 T. This feature is associated
with the high field ordered phase. We found that the fields where
the $M(H)$ kinks occur are the same for $H \parallel [111]$ and
[001] (i.e., no anisotropy). In contrast, Tenya et~al. reported
small but noticeable anisotropy in the location of the boundary of
the high field ordered phase.~\cite{Tenya02} A peak effect (PE) in
the superconducting state that had an onset at $\sim$ 1.25 T and
disappeared at $\sim$ \mbox{0.3 T} below $H_{c2}(T)$ was also
observed (Fig.~\ref{fig:PEH111}). According to recent thermal
conductivity measurements on a \PrOsSb{} single crystal in a
magnetic field,~\cite{Izawa02} there are two distinct
superconducting phases in the $H-T$ plane with two-fold and
four-fold symmetries. The phase boundary between the two-fold and
four-fold symmetry superconducting phases is at \mbox{$H \approx$
0.75 T} for \mbox{$T \approx$ 0.5 K}. The onset field of our PE
occurs at a higher field and for the fields below the PE region no
anomaly is observed in our $M(H)$ curves. It is not clear whether
the PE is related to the two-fold and four-fold symmetry
superconducting phases~\cite{Izawa02} or to pinning from crystal
defects or impurities. However, recent transverse-field muon spin
rotation measurements~\cite{MacLaughlin02} with $H =$ \mbox{200
Oe} indicated that \PrOsSb{} has a nearly isotropic
superconducting energy gap in the superconducting state.

\begin{figure}
 %\begin{center}
  \epsfbox{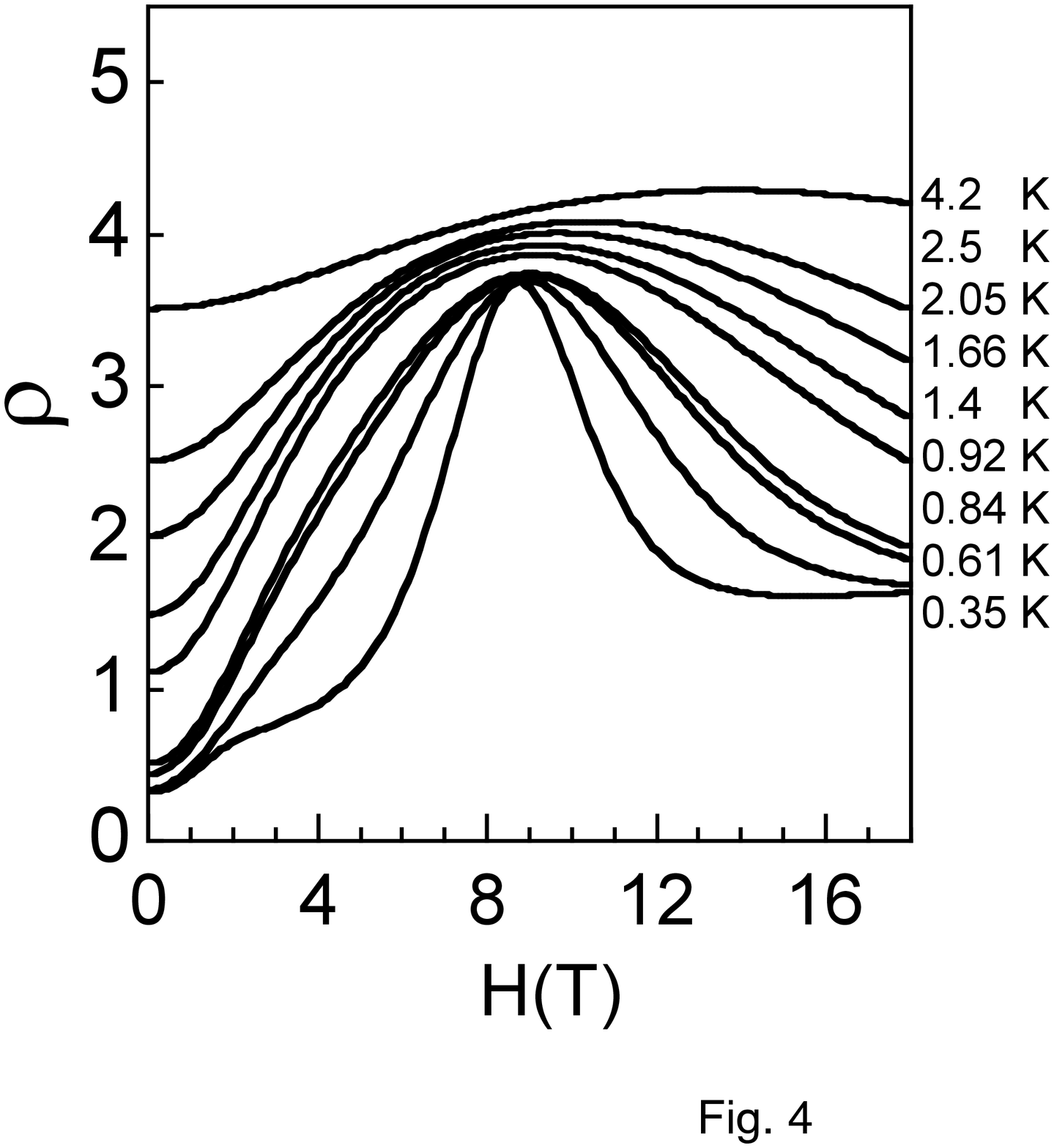}
 %\end{center}
 \caption{Calculated $\rho$ vs $H$ isotherms based on a CEF Hamiltonian including
          equal amounts of the magnetic exchange and aspherical Coulomb
          interactions assuming a $\Gamma_3$ ground state (ref. 17). The dome-shaped
          structure is qualitatively comparable to Fig.~\ref{fig:rhoT&H}(c).}
          %~\cite{Frederick03}
 \label{fig:ACSrhoH}
\end{figure}

%phase diagram
Evidence for a high field ordered phase (HFOP) was derived from
the kinks in magnetoresistivity $\rho(H,T)$ data,~\cite{Ho02}
kinks in magnetization $M(H,T)$ data,~\cite{Tenya02} peaks in
specific heat $C(T,H)$ data,~\cite{Vollmer03} and peaks in thermal
expansion $\alpha(T,H)$ data.~\cite{Oeschler02} The $\rho$ vs $H$
isotherms reveal that $\rho$ is enhanced in the HFOP, over a
linear interpolation of $\rho$ from outside this region
(Fig.\ref{fig:rhoT&H}(c)). From our new set of $\rho(H,T)$
measurements (Fig.\ref{fig:rhoT&H}), the upper boundary of the
HFOP can be determined. The $H-T$ phase diagram, depicting the
superconducting and the HFOP regions, is constructed in
Fig.~\ref{fig:PhaseDiagram}. Since an increase in the magnetic
field would induce mixing of the ground state and the excited
state, the HFOP may be driven by the crossing of the upper level
of the $\Gamma_3$ doublet and the lowest level of the $\Gamma_5$
triplet states at \mbox{$\sim$ 4.5 T} and another crossing between
the lowest levels of the $\Gamma_3$ and $\Gamma_5$ states (change
of the ground state) at \mbox{$\sim$ 10 T}.~\cite{Vollmer03}
Calculations of $\rho(H,T)$ for the case in which a magnetic
exchange interaction and an aspherical Coulomb
interaction~\cite{Fulde72,Anderson74,Fisk77} between the Zeeman
split 4f and conduction electrons has been added to the CEF
Hamiltonian with a $\Gamma_3$ ground state at $H= 0$ T agree
qualitatively with the experimental $\rho$ vs $H$
isotherms.~\cite{Frederick03} The calculated $\rho$ vs $H$
isotherms are plotted in Fig.~\ref{fig:ACSrhoH} and compared with
the experimental $\rho$ vs $H$ isotherms in
Fig.~\ref{fig:rhoT&H}(c). In this calculation, we found that a
$\Gamma_1$ ground state will not produce the same features we
observe in $\rho-H$ measurements. This provides further support
for a $\Gamma_3$ ground state in \PrOsSb{}. In
Fig.~\ref{fig:PhaseDiagram}, the dashed line below \mbox{4.5 T}
and \mbox{2 K} that connects the peaks in $d\rho/dT$ (ref. 6)
%~\cite{Ho02}
and intersects the HFOP is a measure of the Zeeman splitting
between the Pr$^{3+}$ ground state and the first excited state.
The nearly temperature-independent boundary of the HFOP at $\sim$
14.5 T resembles the antiferroquadrupolar ordered phase observed
in PrPb$_3$,~\cite{Tayama01} which also has a $\Gamma_3$ ground
state. It is plausible that the HFOP in \PrOsSb{} has the same
origin. However, the antiferroquadrupolar ordered phase observed
in PrPb$_3$ has strong anisotropy at high fields. We did not
observe such an anisotropy in the magnetization data of \PrOsSb{}
along [001] and [111] below 5.5 T, although the anisotropy at
these fields may not be large enough to resolve. Neutron
scattering experiments~\cite{Lynn} on our powdered samples of
\PrOsSb{} did not detect any signs of an HFOP.
%However, neutron scattering did not show specific feature at high
%fields, that may be due to the sample quality.

\begin{figure}
 %\begin{center}
  \epsfbox{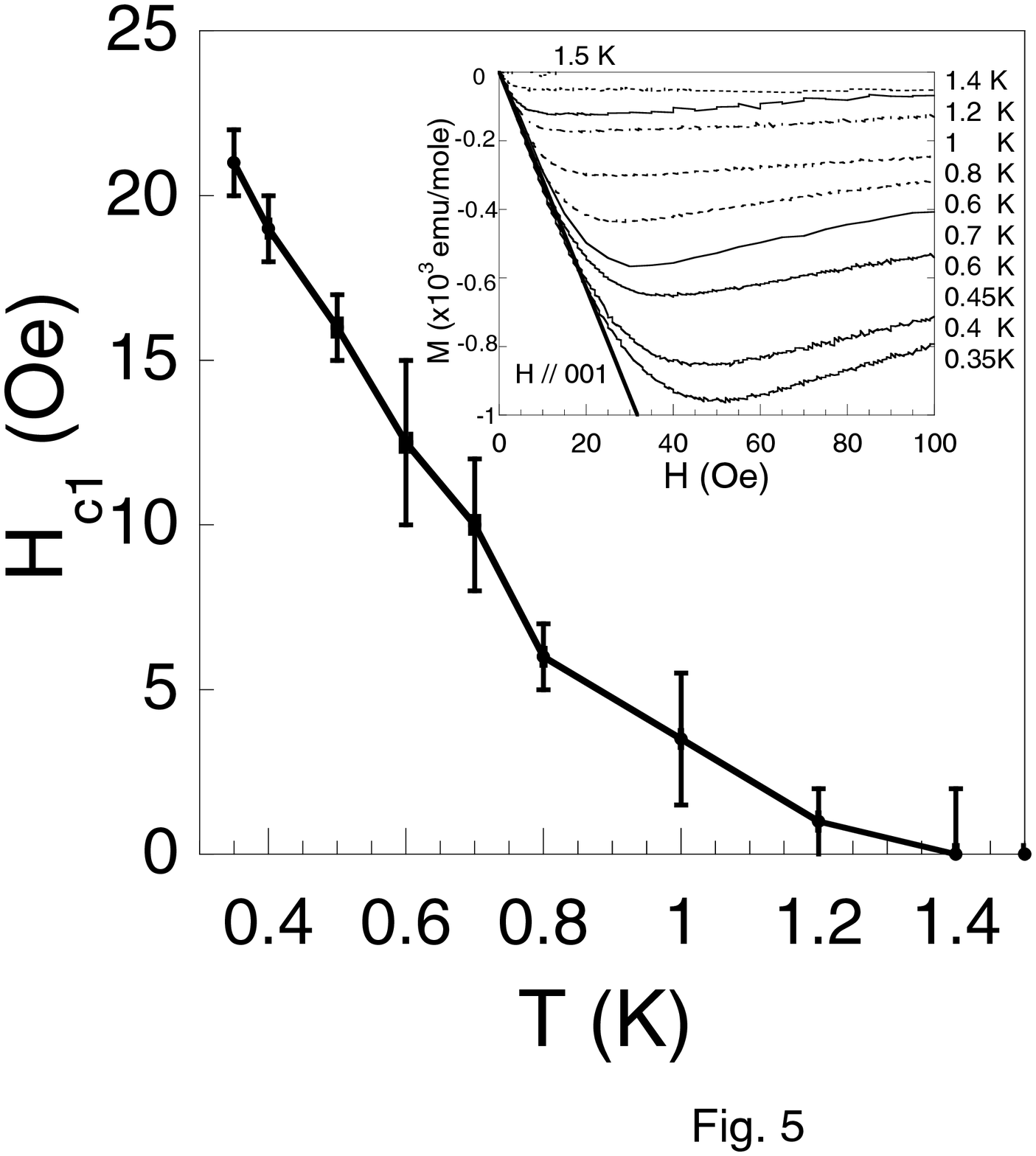}
 %\end{center}
 \caption{Lower critical field $H_{c1}$ vs $T$.
          Inset: $M$ vs $H$ ($\parallel [001]$) below 100 Oe at various temperatures.
          $H_{c1}$ is defined as the field where the curve departs from
          the initial linear region.}
 \label{fig:Hc1H001}
\end{figure}

%SC volume fraction of the sample
%irregular bar shape [~0.086975"(a)]x[~0.0282"(b)]x[~0.02415"(c)] H//c
%c/a~0.278, b/a~0.324. assume c/a=b/a=0.3 for a ellipsoid, n_c=0.4523
Fig.~\ref{fig:Hc1H001} shows the lower critical field $H_{c1}$ vs
$T$ determined for $H \parallel [001]$. $H_{c1}(0)$ is quite
small, $\sim$ \mbox{23 Oe}. The magnetic susceptibility $\chi$ in
the Meissner state was found to be \mbox{-31.5 cm$^3$/mol} (or
-1/[0.613(4$\pi$)]). The demagnetization factor of the sample is
estimated to be $\sim 0.45$,~\cite{Cronemeyer91,PrOsSbDemagFactor}
resulting in an estimate for the superconducting volume fraction
of \mbox{$\sim 74\%$}. The Ginzburg-Landau parameter is estimated
to be $\kappa \approx (H_{c2}/H_{c1})^{1/2} \sim 31$ with the
upper critical field $H_{c2}\sim$ \mbox{2.19 T} and $H_{c1} \sim$
\mbox{23 Oe}. This value is a factor of 10 larger than that
estimated from the relation $\kappa=\lambda/\xi_0\sim 3$, where
the penetration depth $\lambda\sim$ \mbox{344 \AA} was taken from
$\mu$SR measurements~\cite{MacLaughlin02} and the coherence length
$\xi_0\sim$ \mbox{113 \AA} was estimated from the initial slope of
$H_{c2}$.~\cite{Ho02} The discrepancy may arise from error in the
estimates of $H_{c1}$ due to the residual field in the
superconducting magnet or to different qualities of samples used
in the various measurements.

\section{Summary}
In summary, we have reported electrical resistivity and
magnetization measurements on \PrOsSb{} at high magnetic fields
and determined the $H-T$ phase diagram below \mbox{18 T}.

Aspherical Coulomb and magnetic exchange scattering between 4f and
conduction electrons can qualitatively describe the dome shape
features observed in $\rho(H)$ data and support the evidence that
$\Gamma_3$ is the ground state at zero magnetic field.

The high field ordered phase (HFOP) boundary is determined from
kinks in $\rho(H,T)$ and $M(H,T)$ data. The HFOP is confined to a
region on the $H-T$ plane between \mbox{$\sim$ 4.5 T} and
\mbox{14.5 T}, and below \mbox{$\sim$ 1 K}. Measurements of
$M(H,T)$ for $H <$ \mbox{5.5 T} and $\parallel$ [001] and [111]
directions did not exhibit appreciable anisotropy. In analogy with
the behavior of PrPb$_3$, the HFOP may be associated with
antiferroquadrupolar order.

The $M(H,T)$ measurements revealed a small value for the lower
critical field $H_{c1}(0)$ of $\sim$ \mbox{23} Oe, and a value for
the upper critical field $H_{c2}(0)$ of $\sim$ \mbox{2.19 T},
yielding a Ginzburg-Landau parameter $\kappa \sim$ 31. A peak
effect was observed which had an onset at $\sim$ \mbox{1.3 T}
between \mbox{0.35 K} and \mbox{0.7 K} and disappeared at a field
that tracked the $H_{c2}(T)$ curve which had an offset of $\sim$
\mbox{0.3 T}. No evidence in the $M(H,T)$ data was found for a
cross-over between two superconducting phases with different order
parameter symmetry as reported by Izawa et al.~\cite{Izawa02}

\section{Acknowledgement}
This research was supported by the \mbox{U. S.} Department of
Energy under Grant No. DEFG-03-86ER-45230, the \mbox{U. S.}
National Science Foundation under Grant No. DMR-00-72125, and the
NEDO international Joint Research Program, and the work at the
NHMFL Pulsed Field Facility (Los Alamos National Laboratory) was
performed under the auspices of the NSF, the State of Florida and
the US Department of Energy.

% Create the reference section using BibTeX:
%\bibliography{CeRu2MrlxPE}

\begin{thebibliography}{10}
\bibitem{PrOsSbDemagFactor}
The PrOs$_4$Sb$_{12}$ single crystal used in the magetization
meausrements had an irregular bar shape with $a \times b \times c
\approx$ 2.2 mm $\times$ 0.72 mm $\times$ 0.61 mm and $H \parallel
c$. It was assumed ro be equivalent to an ellipsoid with $c/a \sim
b/a \sim 0.3$.
\end{thebibliography}

\newpage

% fig.1
\epsfxsize=500pt \epsfysize=550pt

% fig.2
\epsfxsize=500pt \epsfysize=550pt

% fig.3
\epsfxsize=500pt \epsfysize=550pt

% fig.4
\epsfxsize=500pt \epsfysize=550pt

% fig.5
\epsfxsize=500pt \epsfysize=550pt

\end{document}